# Scalable universal tunable virtual-wavevector spatial frequency shift (TVSFS) super-resolution imaging

Mingwei Tang,[1] Yubing Han,[1] Dehao Ye,[1] Qianwei Zhang,[1] Chenlei Pang,[1,2] Xiaowei Liu,[1,2] Weidong Shen,[1] Yaoguang Ma,[1] Clemens F. Kaminski,[3] Xu Liu,[1,2,4,*] and Qing Yang[1,2,4,*]

[1] State Key Laboratory of Modern Optical Instrumentation, College of Optical Science and Engineering, International Research Center for Advanced Photonics, Zhejiang University, Hangzhou 310027, China
[2] Research Center for Intelligent Sensing, Zhejiang Lab, Hangzhou 311100, China.
[3] Department of Chemical Engineering and Biotechnology, University of Cambridge, Cambridge, CB30AS, United Kingdom
[4] Collaborative Innovation Center of Extreme Optics, Shanxi University, Taiyuan, 030006, China
*Corresponding authors: liuxu@zju.edu.cn, qingyang@zju.edu.cn

## Abstract

Developing a chip-based super-resolution imaging technique with large field-of-view (FOV), deep subwavelength resolution, and compatibility for both fluorescent and non-fluorescent samples is desired for material science, biomedicine, and life researches, etc. Previous on-chip super-resolution methods focus on either fluorescent or non-fluorescent imaging, putting an urgent requirement on the general imaging technique compatible with both of them. Here, we introduce a universal super-resolution imaging method based on tunable virtual-wavevector spatial frequency shift (TVSFS), realizing both labeled and label-free super-resolution imaging on a single delicately fabricated scalable photonic chip. Theoretically, with TVSFS, the diffraction limit of a linear optical system can be overcome, and the resolution can be improved more than three times, which is the limitation for most super-resolution imaging based on spatial frequency engineering. Diffractive units were fabricated on the chip's surface to provide a wavevector-variable evanescent wave illumination and induce tunable deep SFS in the samples' Fourier space. A resolution of $\frac{\lambda}{4.7}$ for label-free sample and $\frac{\lambda}{7.1}$ for labeled sample with a large FOV could be achieved with a CMOS-compatible process on a GaP chip. The large FOV, high-compatibility, and high-integration TVSFS chip may advance the fields like cell engineering, precision inspection in the industry, chemical research, etc.

## 1. Introduction

The spatial resolution of conventional microscopy is limited by optical diffraction. The advent of fluorescence-based super-resolution microscopy opens the way for far-field detection in discerning nanometer-scale details, such as stimulated emission depletion microscopy(STED)[1,2], single-molecule localization

microscopy(SMLM)[3], and MINFLUX[4,5]. Such methods break the diffraction limit in the spatial domain by shrinking the point-spread function of the optical system. Although the above methods provide a significant improvement in achieving spatial resolution, they also suffer from the requirement of special labeling.

The reason for the diffraction limit is that the high spatial-frequency information locates within the near-field region of the specimen's surface and can not be detected by the far-field objective. Spatial frequency shift (SFS) method[6-8] can break the diffraction limit by manipulating the spatial frequency domain instead of the spatial domain. The SFS methods use multiple light waves with large-transverse wavevectors to illuminate the sample and collect the spatial-frequency shifted wide-field images using conventional objectives in the far-field. By reconstructing in the Fourier space, the detection spatial-frequency aperture (that is, the transfer function) can be expanded. Therefore, it can provide high-speed, high-resolution, and wide-field imaging. Compared with STED and SMLM methods, which rely on fluorescence with specific characters, the SFS method is a universal method compatible with both label-free and labeled imaging.

Conventional SFS microscopy examples are structured illumination microscopy (SIM)[9] for labeled imaging and Fourier ptychographic microscopy (FPM)[10,11] for label-free imaging. Both of them use real-wavevector free-space light for illumination, making the extension in Fourier space limited by the refractive index of imaging immersion medium. By introducing large virtual-wavevector evanescent waves for illumination, the resolution of SFS imaging can reach the subwavelength scale and overcome the diffraction limit. Furthermore, existing SFS imaging methods are only compatible with label-free[7,8,12-14] or labeled imaging[15-18] for lacking a universal wavevector tuning method for both samples and the flexible scheme switching between them.

Here, we propose a universal tunable virtual-wavevector spatial frequency shift (TVSFS) method, which can break the resolution limit significantly using the virtual-wavevector illumination for both label-free and labeled imaging. The proposed TVSFS method is implemented on a GaP photonic chip with fabricated micro-/nanostructures, which is CMOS compatible and can serve as a multifunction platform for cell stimulation, imaging, and selection. The delicately designed nanogratings control the wavevector direction and magnitudes precisely to realize multilevel tuning for a wide coverage of the Fourier spectrum.

## TVSFS imaging on a chip

Chip-based super-resolution has gained popularity in recent years for its promise in simplifying the super-resolution implementation and reducing maintenance costs[15,19-21]. The lateral-wavevector and light propagation loss of the super-resolution chip should be carefully weighed to obtain a large-wavevector and large-FOV illumination[22-24]. Plasmonic structures generating surface plasmons promise to provide illumination with a large lateral-wavevector but at the cost of a non-negligible ohmic loss[25], thus decreasing the continuous FOV. Dielectric materials, like $SiO_2$, $Al_2O_3$, can propagate light to a longer distance than plasmons based on metals but can only provide a limited lateral wavevector. Here, we choose a transparent dielectric substrate with a high refractive index[26,27], combining with fabricated micro-/nanostructures and a special stimulation method to realize both large-wavevector and large-FOV illumination.

Wavevector-tunable evanescent waves are generated on the surface of a high-refractive-index chip (GaP in this case) as the illumination source. To realize deep-subwavelength resolution, the magnitude of the illumination wavevector should be prominent. However, when the SFS wavevector is above double the NA of the collection objective, a missing gap will exist in the Fourier space, leading to artifacts in the final image and even failing the image reconstruction[18](Fig.1a). Wavevector tunability is one way of solving this problem. The previous wavevector tuning method either uses the wavelength tuning for label-free imaging[7] or multi-angle tuning[15,18] for labeled imaging. Here we develop a wavevector tuning method to make the photonic chip compatible with both fluorescent and non-fluorescent super-resolution imaging. The wavevector tunability is achieved using period-varied subwavelength gratings fabricated at the surface's predefined locations. On the other side of the chip, samples are distributed on the overlapping region of different direction illumination. Both label-free and labeled subwavelength resolution imaging can be realized, depending on the switching between single-beam illumination (Fig.1b) and double-beam illumination (Fig.1d). The arrows in different colors represent wavevectors with different magnitudes for illumination. To fully understand the principle of the resolution enhancement in the TVSFS method, we need to look into the Fourier space, as indicated in Fig.1c,e. The Fourier space of TVSFS is composed of various circles representing the transfer function of the objective lens. At the center exists the low-pass spectrum of the conventional microscope. The Fourier space can be enlarged and filled to achieve isotropic super-resolution imaging, in which the circles with different colors represent the spectrum with various spatial frequency shift directions and magnitudes. For TVSFS label-free imaging, only one-beam evanescent illumination is needed at every image acquisition. In the TVSFS labeled imaging, two-beam evanescent illuminations encounter at the center of the chip's top surface and interfere with each other to form periodic structured light for illumination (Fig. 1d). An array of gratings with various orientations enables multi illumination angles at the overlapping area. By controlling the phases of the interference arms, three raw images (three phases per rotational angle) are required for one spatial frequency shift magnitude. The label-free and labeled super-resolution imaging can use the same chip with identical gratings, depending on the scheme of evanescent wave stimulation.

In our chip design, the multilevel tuning is obtained by using gratings with different periods and directions. The gratings on the chip's surface have a corresponding relationship with the TVSFS shift wavevectors in the Fourier space (Fig. 1b, d). For gratings with the same period, different directions are needed along a circle, which centers at the sample region on the other surface of the chip. The radius of the circle ($r$) relates to the thickness of the chip substrate ($T$) and the angle of light deflected from the gratings ($\theta$) as: $r = T \cdot tan(\theta)$.

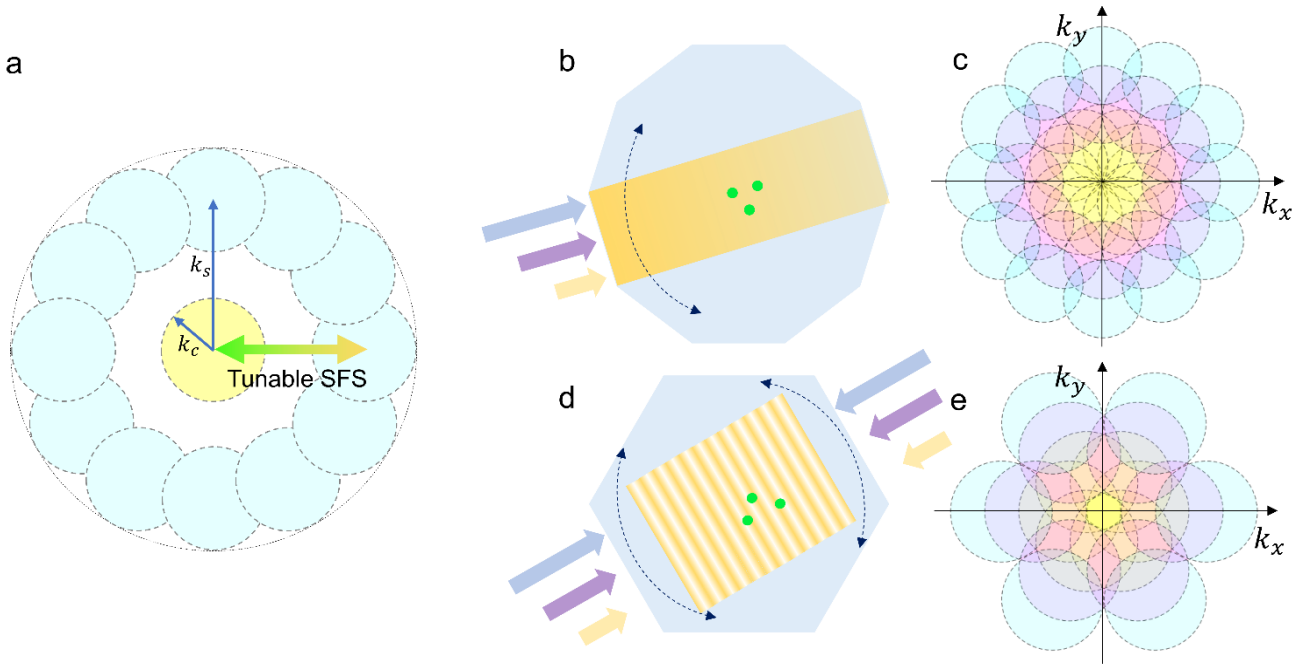

**Fig.1 | Physical Scheme of chip-based TVSFS super-resolution imaging. a**, The importance of multilevel tuning for SFS imaging. **b**, The single-beam illumination for label-free imaging. **c**, The Fourier space of label-free TVSFS imaging. **d**, The two-beam illumination for labeled imaging. **e**, The Fourier space of labeled TVSFS imaging.

The resolution of the TVSFS method is basically determined by the refractive index ($n_{PC}$) of the photonic chip. For every image acquisition process of a single frame, the detected spatial spectrum can be described as:

$$F_d = F_o(\vec{k} - \vec{k_s}) \cdot TF(\vec{k}) \tag{1}$$

where $F_o$ is the Fourier transform of the object, $\vec{k_s}$ represents the spatial-frequency shift vector of the acquisition process, $TF(\vec{k})$ is the transfer function of the objective lens, which is bound by the NA of the objective lens. The resolution is determined by both the aperture of the system $NA \cdot \frac{2\pi}{\lambda_{em}}$ and the spatial-frequency shift magnitude that can be provided by the evanescent illumination module, which is $n_{PC} \cdot sin\theta \cdot \frac{2\pi}{\lambda_{ex}}$, where $\lambda_{em}$ and $\lambda_{ex}$ are the wavelength of the emission and exciting light. Therefore, the theoretical formula describing the resolution limitation of TVSFS takes the form of:

$$\Delta_{xy} = \frac{\lambda_{em}}{(NA + n_{PC} \cdot sin\theta)} \tag{2}$$

for label-free imaging;

$$\Delta_{xy} = \frac{\lambda_{em}}{2\left(NA + \frac{n_{PC} \cdot sin\theta \cdot \lambda_{em}}{\lambda_{ex}}\right)} \tag{3}$$

for labeled imaging.

If the spatial-frequency shift magnitude is larger than the NA, the resolution exceeds conventional wide-field and SIM microscopy. In the TVSFS method, the resolution can be scaled by changing the spatial frequency shift magnitude using different periods of coupling gratings through changing $sin\theta$ (Fig. 1b, d). For a deep spatial frequency shift, the varied spatial frequency shift is non-trivial for filling the spatial frequency space. Therefore, gratings with different periods should be fabricated and made sure the spectrum overlapping between two successive illuminations exceeds 35%[28]. Here we use a photonic chip with a high refractive index ($n_{PC} = 3.43$ at the wavelength of 561 nm), far exceeding the highest NA of the existing objective lens (NA ≈ 1.7). The attainable resolution can be as low as 55 nm for labeled imaging and 110 for label-free imaging. We design gratings with three periods to achieve the multilevel SFSs.

The FOV of TVSFS is determined by the illumination size of the evanescent wave, which is set by the size of fabricated gratings. In the present work, we have achieved $40 \times 40$ μm² gratings by etching with focused ion beam (FIB), but the size can be over $100 \times 100$ μm² in the further attempt. In the TVSFS method, the illumination module and the collection objective lens are decoupled; therefore, we can use a low-NA objective lens combined with a large-wavevector illumination module to achieve a high resolution and a large FOV simultaneously.

## 2. Results and discussion

To experimentally demonstrate the TVSFS super-resolution imaging, we developed a fabrication process for the scalable chip with various controllable illumination wavevectors. We used FIB to fabricate the gratings with different periods at the predesigned position. The illumination wavevectors are scalable by modifying the design parameters, such as the thickness of the substrate, wavelength of the illumination source, and period of the gratings. In this work, we demonstrate the 2-inch wafer-scale TVSFS chip using the standard CMOS fabrication process. More than 100 chip arrays can be fabricated in a batch and divided for use. Background light depression is the crucial step to make a successful TVSFS imaging since the coupling of light with the gratings will introduce many stray lights. To achieve this, we have developed a light-blocking design by using dual surface lithography followed by metal depositing. Besides providing an unlimited FOV, the transparent property of GaP substrate in the visible wavelength also makes the dual surface lithography process possible with a standard lithography machine.

## TVSFS label-free imaging

We first investigate the label-free imaging capability of our TVSFS chip. The necessity of multilevel tuning of illumination wavevector is theoretically simulated. A 'ZJU' eagle logo pattern with a double-line-profile is chosen as the sample, of which the line-to-line distance was set as 215 nm. The spatial distribution of this pattern under multilevel wavevectors illumination is presented in Fig.2a, where $K_0 \sim K_3$ represent wavevectors with increased magnitudes. The wavevectors are chosen with enough overlap in the Fourier space between each other to ensure a good convergence in the reconstruction process. The reconstructed image with a whole spectrum is shown in Fig.2g, which covers the low-SFS, middle-SFS, and high-SFS spectrum band. Compared with the vertical illumination ( $K_0$ ) with a 0.85-NA objective (Fig.2b), whole spectrum reconstruction proves the perfect resolving ability of the double-line details. Fig.2c-f show the reconstructed result in various cases, with partial spectrum band omitted. Without multilevel tuning of the illumination wavevector, the final image either lacks a deeper resolution (Fig.2c,d) or exhibits the false reconstruction (Fig.2e,f).

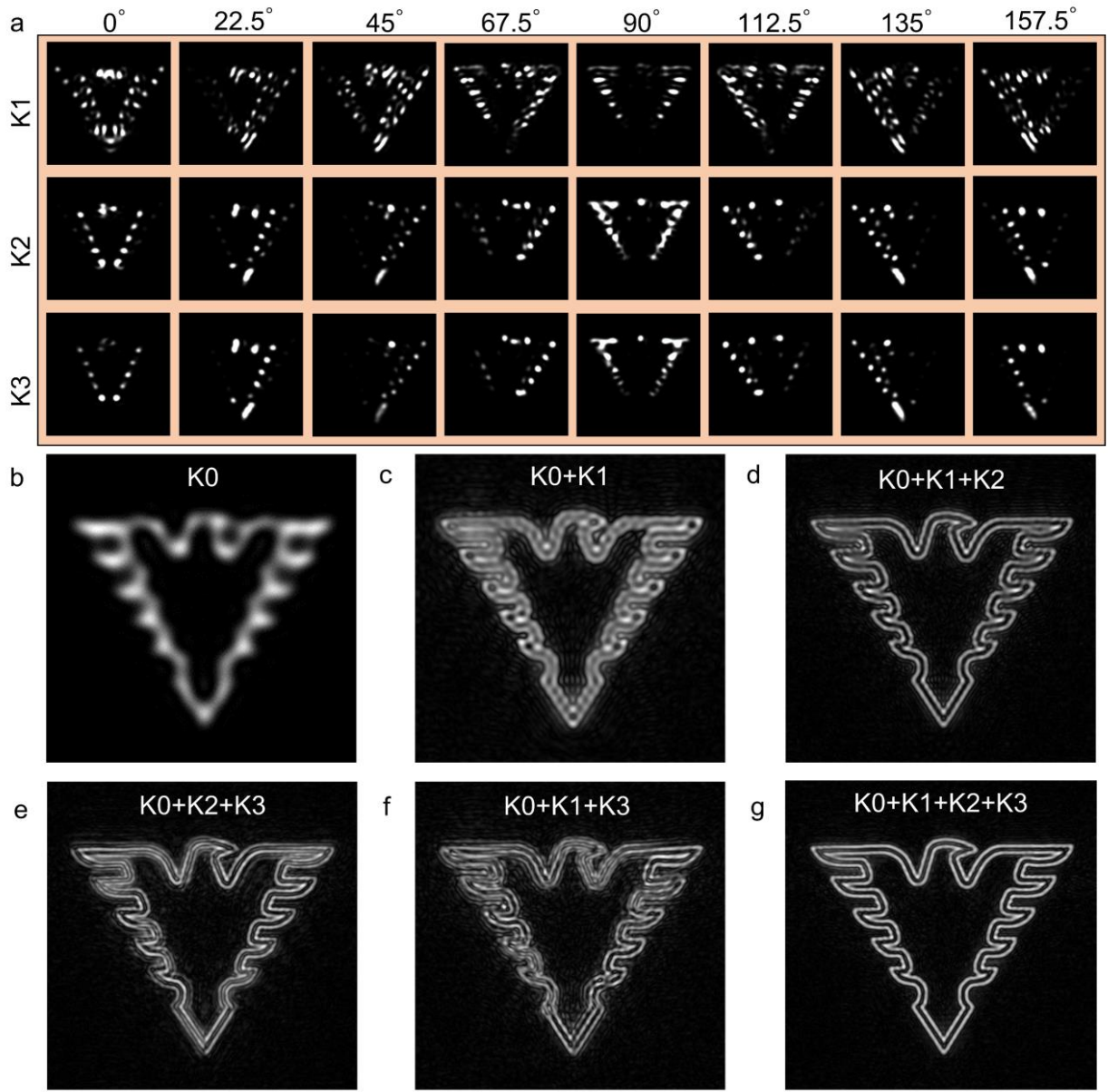

**Fig.2 | Simulation of TVSFS label-free imaging. a**, The different part of the spatial spectrum of a 'ZJU' eagle logo pattern, represented in the spatial domain. **b**, The low-pass spatial spectrum by the conventional objective. **c-g**, The reconstructed result of TVSFS containing partial spatial spectrum.

In the experiment, the 'ZJU' eagle logo pattern was etched on the GaP wafer surface using FIB, with the same parameters for comparison, as indicated in the SEM image (Fig.3b). We used a 660-nm wavelength laser diode for illumination. The laser was spatially filtered before coupling into the photonic chip through the gratings. The speckle noise is severe in the raw images using the laser source directly. Previous coherent imaging methods usually reduce the speckle noise using light sources with less spatial or spectral coherence.[7,8,10] In our case, this will decrease the image contrast for the introduction of extra background light. Here we use the wavelength-tunable laser diode modulated by a zigzag wave, which can efficiently depress the speckle noise and improve the signal-to-noise ratio (SNR) of the collected image. The experimentally acquired raw images have a good match with that obtained by our theoretical simulation using the same parameters (Fig.2a). Some mismatch may be caused by the defect in the sample fabrication. The uniformity of the grating fabrication and refractive index of the photonic substrate is also crucial for obtaining a perfect raw image. With our spatial spectrum splicing algorithm, the final high-resolution image of the sample can be obtained (Fig.3b). Compared with that directly acquired by wide-field illumination, the TVSFS label-free imaging has a 4.5 × resolution enhancement (see the Fourier spectra of bottom-right corners of Fig.3b-d). The line comparison between SEM and TVSFS method (Fig.3e) shows an excellent match, demonstrating our method's authenticity. The line PSF of the TVSFS label-free method is 81 nm, demonstrating a double-line resolving ability of around 162 nm. To further push the resolution of our method, a shorter wavelength (561 nm) and 1.49-NA objective are used. The imaging of a 4-line etched sample validates our TVSFS label-free has a resolution down to 120 nm.

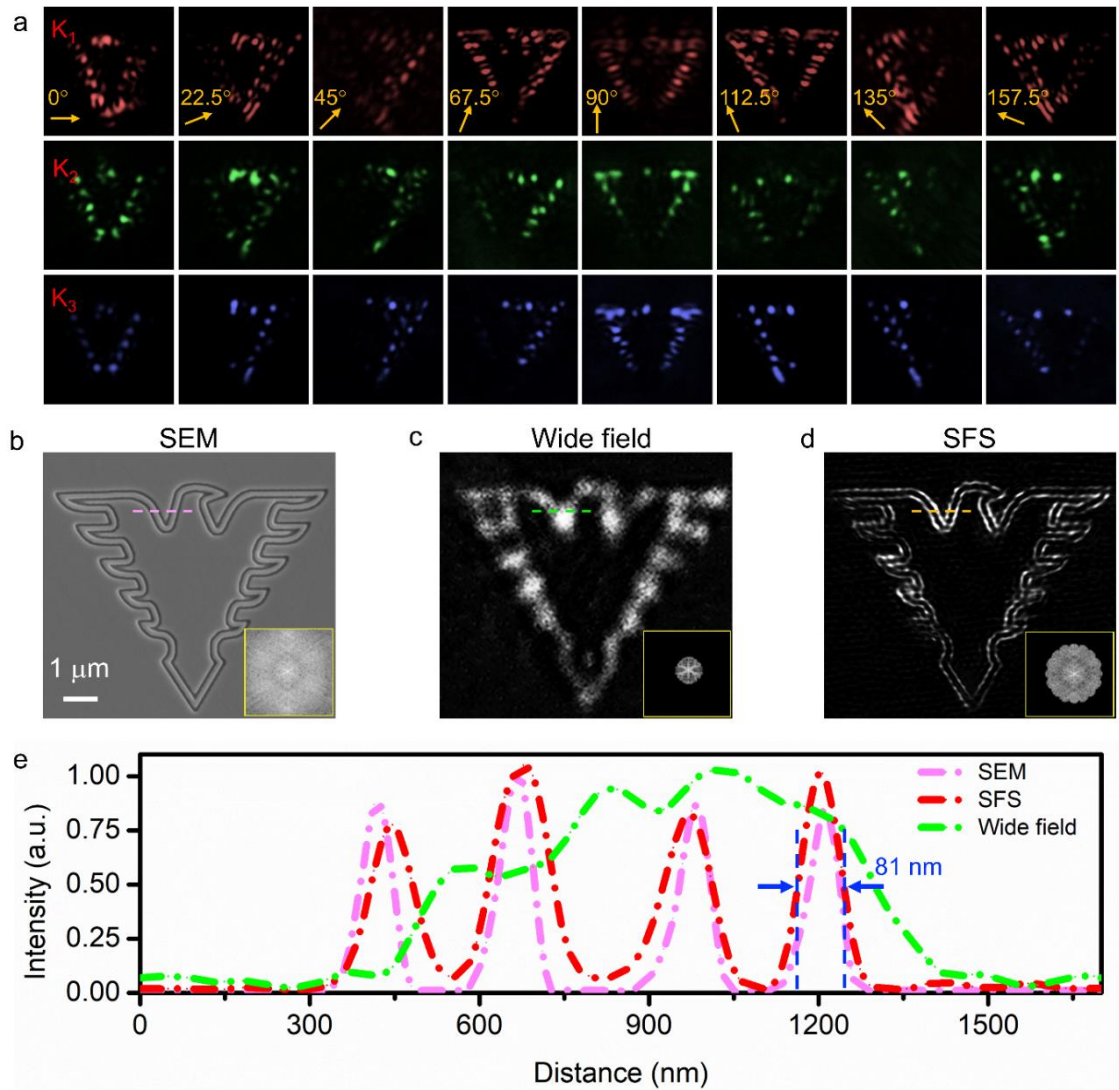

**Fig.3 | On-chip TVSFS label-free imaging of etched 'ZJU' eagle logo patterns. a**, TVSFS label-free raw images acquired under eight directions and three wavevectors. Different colors represent wavevector with various magnitudes. **b-d**, The 'ZJU' eagle logo image taken with SEM (**b**), wide-field under vertical illumination (**c**), and TVSFS label-free method (**d**). The Fourier spectra are shown in the bottom-right corner of (**b-d**). **e**, Line profiles, taken at the position indicated with a dashed line in **b-d**, showing that line details that are not resolved in the wide-field image( dashed green line) are clearly resolved in the TVSFS image (dashed red line) and has a good match with that of SEM image (dashed magenta line). The intensity profile in SEM and wide-field are inverted to better show the relationship with TVSFS imaging, which is a dark-field image in this case.

**TVSFS labeled imaging**

Unlike label-free imaging using uniform evanescent waves for coherent illumination, the periodic patterns formed by the interference between two counterpropagating evanescent waves are applied for fluorescence illumination.[29] To experimentally demonstrate the method, grating couples are stimulated to obtain two counterpropagating evanescent waves. The grating period can be changed to make the spatial frequency shift values adjustable. Chips with an octagonal sample region and three periods of gratings (each period has eight gratings with equally distributed azimuthal angle) were fabricated. The gratings are distributed at the predefined position to make sure the light with wavevectors of $0.9k_0$, $1.6k_0$, and $2.4k_0$ ($k_0 = \frac{2\pi}{\lambda}$ is the wavenumber in a vacuum) propagate to an overlapped region at the sample region. A water-immersion objective lens with NA=1.1 was used to collect the TVSFS raw images. To control the light splitting and phase shifting (three phases per orientation per SFS magnitude) between two light paths, we use a spatial light modulator (SLM) to modulate the laser. For isotropic resolution, the SLM pattern needs to be rotated to have four equally spaced orientations. This only describes the resolution of one spatial frequency shift value; however, to achieve higher resolution, the distance between the double beams should be adjusted to fit the gratings with smaller periods (distributed at the margin of the photonic chip). Here we use the 40-nm fluorescent bead samples to demonstrate the TVSFS labeled imaging. Samples are illuminated by period patterns with various fringe spacings and reconstructed step by step incorporating higher spatial frequency shift magnitudes. The images show the reconstructed images with an improved resolution by taking the larger illumination wavevector into account (Fig. 4c-f). By drawing a line profile (Fig. 4g) across the intensity distribution marked with a white dash line in Fig. 4c-f, what appears as a continuous distribution in the diffraction-limited image (Fig. 4c) clearly shows the details in the TVSFS reconstruction. The intensity profile has a better contrast when the spatial frequency shift value increases from $0.9k_0$ to $1.6k_0$. When the spatial frequency shift value reaches $2.4k_0$, the line profile shows that the two beads are separated by 93 nm and resolved.

Finally, the feasibility of using the on-chip TVSFS platform to image biological specimens was also demonstrated. Human bone osteosarcoma epithelial cells (U2OS) were deposited onto the TVSFS chip surface following the established labeled method. These cells were labeled for actin filaments using Alexa Fluor 647 Phalloidin. The 1.49-NA objective lens combining a chip with a spatial frequency shift value of $1.48k_0$ was used. The aberration of illumination patterns has an important role in the final reconstruction. The illumination patterns along one direction with three-phase shifts using a low illumination intensity were demonstrated in Fig. 5a, which shows nearly no defect. Besides, the line profile of the position marked with a white box in Fig. 5a proves an even phase shift with ±120°(Fig. 5b). The obtained diffraction-limited image and TVSFS reconstruction of the actin filaments are exhibited in Fig. 5c and Fig. 5d. In the enlarged images in Fig. 5e,f, a clear improvement in resolution is observed from the diffraction-limited image (Fig. 5e) to the TVSFS image (Fig. 5f). Fig. 5g presents a line profile across the actin filaments; the TVSFS image visualizes the separation of actin filaments, which is not resolved in the diffraction-limited image.

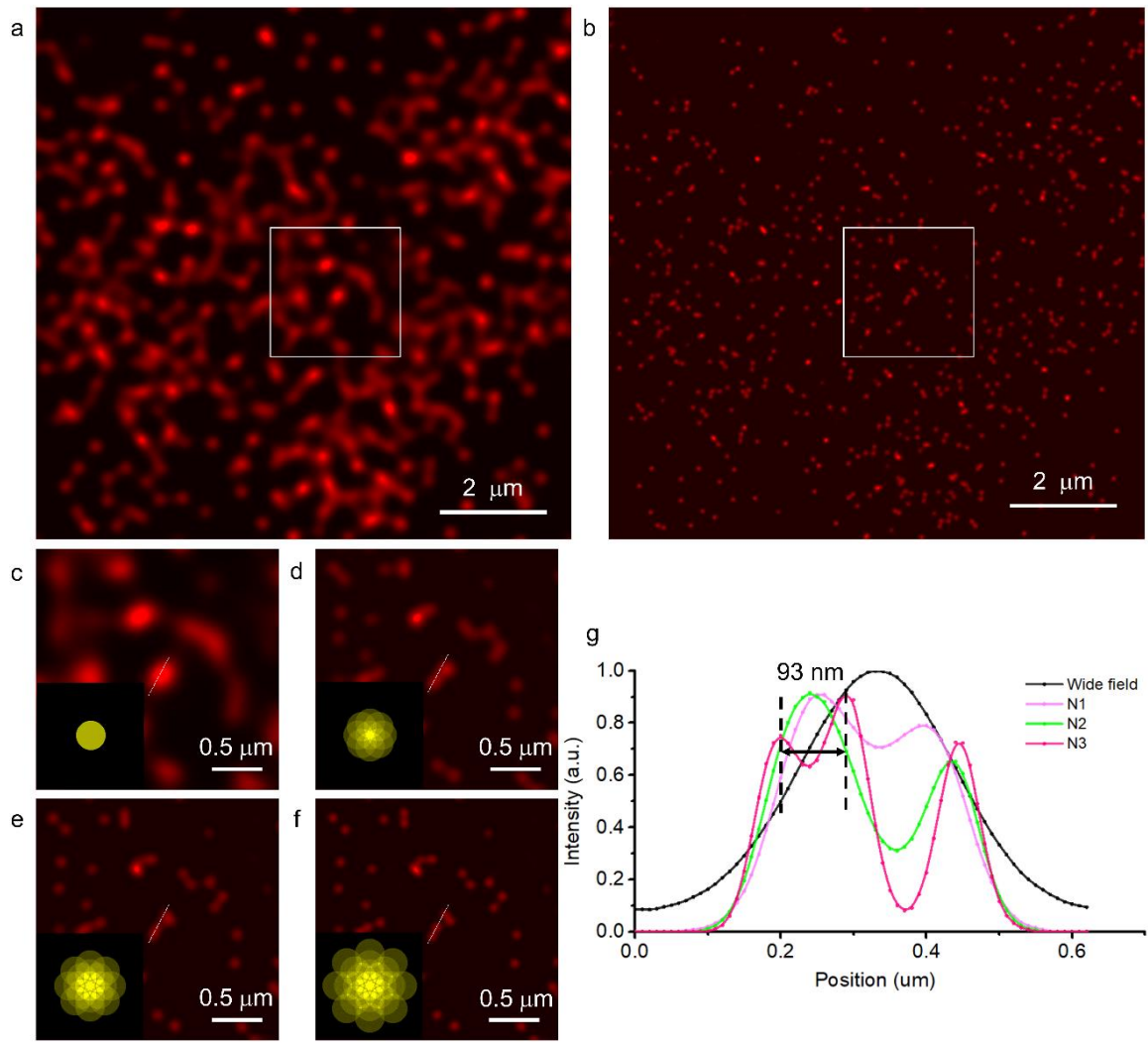

**Fig.4 | Resolving emitters with TVSFS labeled method. a,b**, A 40-nm fluorescent bead sample with diffraction-limited resolution (**a**) and the TVSFS reconstruction of the same region (**b**). **c-f**, Enlarged views of the areas indicated with a white box in **b**, where (**c**) is the wide-field image,(**d-f**) show the images reconstructed by TVSFS method with maximal spatial frequency shift values of $0.9k_0$, $1.6k_0$, and $2.4k_0$. The down left corners show the corresponding spatial spectra. **g**, The white dash lines in **c-f** indicate the position of the line profile in (**g**). Two beads located 93 nm apart are resolved using TVSFS labeled method with a spatial frequency shift value of $2.4k_0$, but not in the diffraction-limited image. The excitation/emission wavelengths are 639 nm/661 nm.

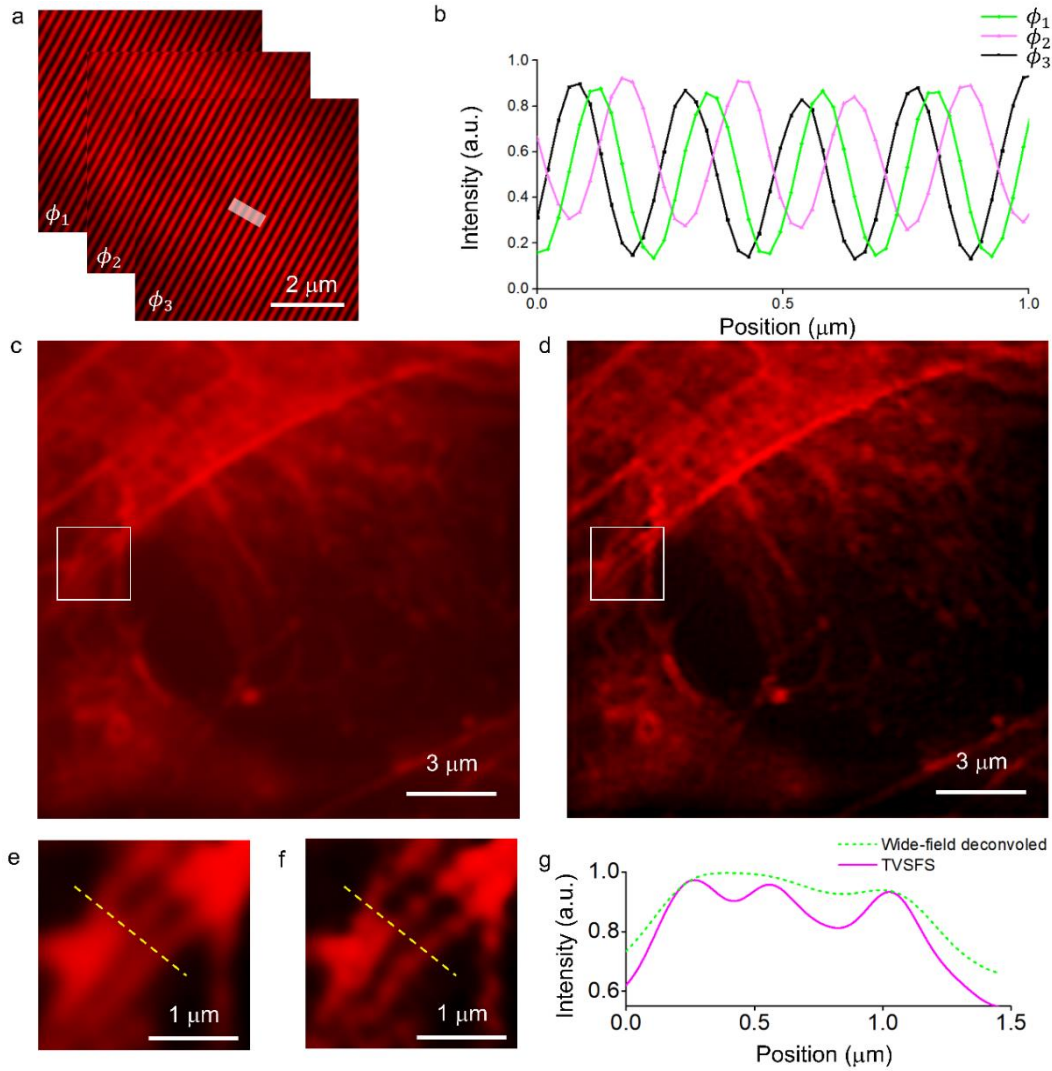

**Fig.5 | Experimental demonstration of TVSFS labeled method. a**, Illumination patterns of TVSFS labeled imaging generated by the photonic chip. The counterpropagating light was designed to have a wavevector of $1.48k_0$. We used a 1.49-NA objective lens to confirm the interference pattern. The polarization direction was changed to maximize the pattern contrast. **b**, Intensity along the line profile in **a** shows an interference pattern with a period of 215 nm, which indicates the evanescent illumination wave has a refractive index of 1.48. **c**, The wide-field fluorescent image of the cell. **d**, A TVSFS image, showing clear resolution enhancement compared to **c**. **e-f**, The enlarged images of the white box in **c-d**. **g**, the line profile in **e-f**.

## 3. Conclusion

In summary, we have demonstrated a universal super-resolution imaging method based on the tunable virtual-wavevector spatial frequency shift effect, which we have termed as 'tunable virtual-wavevector spatial frequency shift' (TVSFS). The TVSFS method can be intrinsically combined with a high-refractive-index photonic chip to provide large-FOV, deep super-resolution imaging. The theoretical modeling of the TVSFS principle, complemented and confirmed by our experimental work, demonstrates the potential of the proposed TVSFS chip to be used for both labeled and label-free samples. The TVSFS photonic chip will open avenues for universal super-resolution imaging with user-defined illumination wavevectors. The parameters of our chip can be easily retrofitted, allowing standard microscopy to acquire super-resolution imaging. Large array chips can be produced in a wafer, appropriate for large production and reducing cost.

The relatively new field of the TVSFS super-resolution method would benefit from future improvements. In this study, excessive data was used to achieve a good reconstruction effect, which will add to the chip fabrication cost and slow down the imaging speed. With the development of better reconstruction methods such as adopting the deep-learning method[30,31], we may further reduce the number of spatial frequency shift values and thus decrease the number of fabricated gratings and captured raw images. The advanced fabrication techniques may also boost the development of the TVSFS chip. In this study, the wavevector-variable illumination was achieved using an off-chip laser source; however, the on-chip evanescent light source could be implemented with minor design changes to the device by bonding commercial light chips like LED or VCSEL to provide cheap, large-producible, and microscope-compatible, multi-mode super-resolution imaging.[32,33]. The highly integrated TVSFS chip may be combined with microfluid and find applications in many fields, such as biology, material, and chemical research.

## 4. Materials and Methods

**Experimental setup.**

The setup was based around a custom-built upright microscope fitted with a sCMOS camera (Hamamatsu, ORCA flash, or Prime95B). The microscope is based on a modular commercial system (Thorlabs, CERNA) equipped with Piezo Objective Scanner (Thorlabs, ZFM2020) for adjusting the focus plane of the objective lens and a 3-dimensional stage (Newport, M-462) for coarse alignment of the chip with respect to the objective lens. The lasers are expanded by a 4-f system(L1, L2) to fill the LCOS (Hamamatsu, X13139-01) and then relayed by another 4-f lens (L3, L4). L5 was used to focus the light onto the gratings of the photonics chip. On-chip TVSFS images were acquired using Zeiss ×100/0.85 Air, Olympus ×100/1.1 Water, or Olympus ×100/1.49 Oil objective lens. In label-free imaging, the 660 nm laser diode is modulated by a zigzag wave with a function generator (Stanford DS345) or focusing the 561 nm laser (MGL-FN-561nm) beam through rotating ground glass. For labeled imaging, a 639-nm laser (Changchun New Industries Optoelectronics Technology Company, MSL-FN-639, 300 mW) was used as the excitation source, and a 661 nm bandpass filter (Edmund, 87-755) was used to block the excitation laser. A half-wavelength plate (HWP) was inserted between the laser and the SLM to rotated the polarization state of the light. The quarter wavelength plate (QWP) and liquid crystal retarder were combined to optimize the light polarization incident onto the gratings.

**Sample preparation.**

The GaP chip was washed with ethyl alcohol and ultrapure water several times and sterilized by ultraviolet light for 30 mins. U2OS (human osteosarcoma cell line; ATCC) cells were cultured in McCoy's 5A medium (Thermo Fisher Scientific, Inc.) supplemented with 10% (v/v) Fetal Bovine Serum (Thermo Fisher Scientific, Inc.). Before labeling, the cells were seeded on the GaP chip and maintained at 37°C in a humidified 5% $CO_2$ environment. After overnight incubation, the cells were washed three times with Phosphate buffered saline (PBS; Thermo Fisher Scientific, Inc.), fixed with 4% paraformaldehyde (Electron Microscopy Sciences) for 10 min, and incubated with Alexa Fluor 647 Phalloidin (Thermo Fisher Scientific, Inc.) for 30 min at 37°C according to the manufacture's introductions. The cells were then washed three times with PBS and mounted in ProLong Diamond Antifade Mountant (Thermo Fisher Scientific, Inc.).

A 1 µl drop of 40-nm-diameter fluorescent beads (FluoSpheres, Thermofisher F8789) was dropped in the sample region using a pipette and allowed to dry with nitrogen completely. The chip was then put in a cell culture dish for water immersion imaging.

## Acknowledgment

The authors are grateful to the supports from the National Natural Science Foundation of China (No. 61735017, 61822510, 62020106002, 31901059, 62005250), the Zhejiang Provincial Natural Science of China (No. LR17F050002), the Zhejiang University Education Foundation Global Partnership Fund, the National Key Research and Development Program of China (2018YFE0119000). The authors gratefully acknowledge the support of the Zhejiang University Micro-nano Fabrication Center. The authors also thank Qiulan Liu for help with optical system construction, Wei Wang for help with chip fabrication.

## References

1    Vicidomini, G., Bianchini, P. & Diaspro, A. STED super-resolved microscopy. *Nat. Methods* **15**, 173-182, doi:10.1038/nmeth.4593 (2018).

2    Gottfert, F. *et al.* Coaligned dual-channel STED nanoscopy and molecular diffusion analysis at 20 nm resolution. *Biophys. J.* **105**, L01-03, doi:10.1016/j.bpj.2013.05.029 (2013).

3    Gu, L. *et al.* Molecular resolution imaging by repetitive optical selective exposure. *Nat. Methods* **16**, 1114-1118, doi:10.1038/s41592-019-0544-2 (2019).

4    Balzarotti, F. *et al.* Nanometer resolution imaging and tracking of fluorescent molecules with minimal photon fluxes. *Science* **355**, 606-612, doi:10.1126/science.aak9913 (2017).

5    Gwosch, K. C. *et al.* MINFLUX nanoscopy delivers 3D multicolor nanometer resolution in cells. *Nat. Methods* **17**, 217-224, doi:10.1038/s41592-019-0688-0 (2020).

6    Tang, M. *et al.* Far-Field Superresolution Imaging via Spatial Frequency Modulation. *Laser Photon. Rev.* **14**, 1900011, doi:10.1002/lpor.201900011 (2020).

7    Pang, C. *et al.* On-chip super-resolution imaging with fluorescent polymer films. *Adv. Funct. Mater.* **29**, 1900126, doi:10.1002/adfm.201900126 (2019).

8    Liu, X. W. *et al.* Fluorescent nanowire ring illumination for wide-field far-field subdiffraction imaging. *Phys. Rev. Lett.* **118**, 076101, doi:ARTN 076101
10.1103/PhysRevLett.118.076101 (2017).

9    Heintzmann, R. & Huser, T. Super-resolution structured illumination microscopy. *Chem. Rev.* **117**, 13890-13908, doi:10.1021/acs.chemrev.7b00218 (2017).

10   Zheng, G., Horstmeyer, R. & Yang, C. Wide-field, high-resolution Fourier ptychographic microscopy. *Nat. Photonics* **7**, 739-745, doi:10.1038/nphoton.2013.187 (2013).

11   Zheng, G., Shen, C., Jiang, S., Song, P. & Yang, C. Concept, implementations and applications of Fourier ptychography. *Nat. Rev. Phys.* **3**, 207-223, doi:10.1038/s42254-021-00280-y (2021).

12   Xu, X. *et al.* Si3N4 waveguide platform for label-free super-resolution imaging: simulation and analysis. *J. Phys. D* **52**, 284002, doi:10.1088/1361-6463/ab1c49 (2019).

13   Ströhl, F. *et al.* Super-condenser enables labelfree nanoscopy. *Opt. Exp.* **27**, 25280, doi:10.1364/oe.27.025280 (2019).

14    Ye, D. *et al.* Low loss and omnidirectional Si3N4 waveguide for label-free spatial frequency shift super-resolution imaging. *J. Phys. D* **54**, 315101, doi:10.1088/1361-6463/abfd6e (2021).

15    Helle, Ø. I. *et al.* Structured illumination microscopy using a photonic chip. *Nat. Photonics* **14**, 431-438, doi:10.1038/s41566-020-0620-2 (2020).

16    Liu, X. *et al.* Wide-field 3D nanoscopy on chip through large and tunable spatial-frequency-shift effect. *arXiv: Optics* (2019).

17    Tang, M., Liu, X., Yang, Q. & Liu, X. *Chip-based wide-field 3D nanoscopy through tunable spatial-frequency-shift effect*. Vol. 11549 PA (SPIE, 2020).

18    Liu, X. *et al.* Chip-compatible wide-field 3D nanoscopy through tunable spatial frequency shift effect. *Sci. China-Phys. Mech. Astron.* **64**, 294211, doi:https://doi.org/10.1007/s11433-020-1682-1 (2021).

19    Diekmann, R. *et al.* Chip-based wide field-of-view nanoscopy. *Nat. Photonics* **11**, 322-328, doi:10.1038/nphoton.2017.55 (2017).

20    Archetti, A. *et al.* Waveguide-PAINT offers an open platform for large field-of-view super-resolution imaging. *Nat. Commun.* **10**, 1267, doi:10.1038/s41467-019-09247-1 (2019).

21    Chazot, C. A. C. *et al.* Luminescent surfaces with tailored angular emission for compact dark-field imaging devices. *Nat. Photonics* **14**, 310-315, doi:10.1038/s41566-020-0593-1 (2020).

22    Wei, F. *et al.* Wide field super-resolution surface imaging through plasmonic structured illumination microscopy. *Nano Lett.* **14**, 4634-4639, doi:10.1021/nl501695c (2014).

23    Ponsetto, J. L. *et al.* Experimental demonstration of localized plasmonic structured illumination microscopy. *ACS Nano* **11**, 5344-5350, doi:10.1021/acsnano.7b01158 (2017).

24    Bezryadina, A., Zhao, J., Xia, Y., Zhang, X. & Liu, Z. High spatiotemporal resolution imaging with localized plasmonic structured illumination microscopy. *ACS Nano* **12**, 8248-8254, doi:10.1021/acsnano.8b03477 (2018).

25    Maier, S. A. *Plasmonics: fundamentals and applications*.    (Springer, 2007).

26    Wilson, D. J. *et al.* Integrated gallium phosphide nonlinear photonics. *Nat. Photonics* **14**, 57-62, doi:10.1038/s41566-019-0537-9 (2019).

27    Aspnes, D. E. & Studna, A. A. Dielectric functions and optical parameters of Si, Ge, GaP, GaAs, GaSb, InP, InAs, and InSb from 1.5 to 6.0 eV. *Phys. Rev. B* **27**, 985-1009, doi:10.1103/PhysRevB.27.985 (1983).

28    Dong, S., Bian, Z., Shiradkar, R. & Zheng, G. Sparsely sampled Fourier ptychography. *Opt. Exp.* **22**, 5455-5464, doi:10.1364/OE.22.005455 (2014).

29    Chakrova, N., Heintzmann, R., Rieger, B. & Stallinga, S. Studying different illumination patterns for resolution improvement in fluorescence microscopy. *Opt. Exp.* **23**, 31367-31383, doi:10.1364/OE.23.031367 (2015).

30    Jin, L. *et al.* Deep learning enables structured illumination microscopy with low light levels and enhanced speed. *Nat. Commun.* **11**, 1934, doi:10.1038/s41467-020-15784-x (2020).

31    Nguyen, T., Xue, Y., Li, Y., Tian, L. & Nehmetallah, G. Deep learning approach for Fourier ptychography microscopy. *Opt. Exp.* **26**, 26470-26484, doi:10.1364/OE.26.026470 (2018).

32    Mei, Y. *et al.* Quantum dot vertical-cavity surface-emitting lasers covering the 'green gap'. *Light Sci. Appl.* **6**, e16199, doi:10.1038/lsa.2016.199 (2017).

33    Johnson, K., Hibbs-Brenner, M., Hogan, W. & Dummer, M. Advances in red VCSEL technology. *Adv. Opt. Technol.* **2012**, 1-13, doi:10.1155/2012/569379 (2012).